# Electrical switching dynamics and broadband microwave characteristics of VO$_2$ RF devices

Sieu D. Ha[1*], You Zhou[1], Christopher J. Fisher[2], Shriram Ramanathan[1], and Jacob P. Treadway[2]

[*]Corresponding author: sdha@seas.harvard.edu

[1]School of Engineering and Applied Sciences, Harvard University, Cambridge, MA 02138

[2]Draper Laboratory, Cambridge, MA 02139

**Abstract**

Vanadium dioxide is a correlated electron system that features a metal-insulator phase transition (MIT) above room temperature and is of interest in high speed switching devices. Here, we integrate VO$_2$ into two-terminal coplanar waveguides and demonstrate a large resistance modulation of the same magnitude (>$10^3$) in both electrically (*i.e.* by bias voltage, referred to as E-MIT) and thermally (T-MIT) driven transitions. We examine transient switching characteristics of the E-MIT and observe two distinguishable time scales for switching. We find an abrupt jump in conductivity with a rise time of the order of 10 ns followed by an oscillatory damping to steady state on the order of several μs. We characterize the RF power response in the On state and find that high RF input power drives VO$_2$ further into the metallic phase, indicating that electromagnetic radiation-switching of the phase transition may be possible. We measure *S*-parameter RF properties up to 13.5 GHz. Insertion loss is markedly flat at 2.95 dB across the frequency range in the On state and sufficient isolation of over 25 dB is observed in the Off state. We are able to simulate the RF response accurately using both lumped element and 3D electromagnetic models. Extrapolation of our results suggests that optimizing device geometry can reduce insertion loss further and maintain broadband flatness up to 40 GHz.

# I. Introduction

Correlated electron materials have exotic properties that arise from strong interactions between charge, spin, lattice, and orbital degrees of freedom. Vanadium dioxide ($VO_2$) is one such system that exhibits a sharp metal-insulator phase transition (MIT) as a function of temperature at ∼67 °C in bulk crystals. Coinciding with the electronic transition is a structural transformation from monoclinic in the room temperature insulating phase to rutile in the high temperature metallic phase. The resistivity modulation can be several orders of magnitude and can also be triggered by external stimuli such as an applied voltage or optical excitation. The large resistivity change is of growing interest in emerging solid state device concepts.

Triggering the MIT by external electrical excitation (E-MIT) through an applied voltage or current in two-terminal devices is of particular interest for electronic devices such as memory and for communications such as radar components. Important metrics for electronic devices utilizing the E-MIT include the magnitude of the resistance change ($\Delta R_E = R(Off)/R(On)$) and the speed of the transition. The E-MIT and the associated resistance change $\Delta R_E$ can be compared with the temperature-driven phase transition (T-MIT) and its associated resistance change $\Delta R_T$ (defined below). $\Delta R_T$ likely represents the maximum resistance change obtainable by any excitation, corresponding to switching of the entire material volume. There have been several reports of E-MIT in the literature, but all show $\Delta R_E$ approximately one order of magnitude lower than $\Delta R_T$, or lower.[1-9] The observation of $\Delta R_E < \Delta R_T$ is likely due to incomplete film switching by electrical stimulus, perhaps through preferential conduction through limited switched portions of the material. Here, we integrate $VO_2$ thin films into two-terminal coplanar waveguides and obtain $\Delta R_E \approx \Delta R_T > 10^3$, to the best of our knowledge the largest resistance change in E-MIT of $VO_2$ observed to date and likely evidence for full switching of the VO$_2$ volume by electrical excitation. Furthermore, we examine transient switching characteristics in such structures and find that the E-MIT may involve multiple time scales and mechanisms. We investigate the corresponding RF *S*-parameter properties up to 13.5 GHz and power handling capabilities for possible RF applications in high frequency filters, modulators, and active microwave device components.[3,5,10-16] We show that high RF input power can reduce the resistance of VO$_2$ similarly as applied DC voltage. We compare transient and RF properties with published results on VO$_2$ and present lumped circuit element and 3D electromagnetic simulations that model the RF properties quite well. Such small signal models may be useful for future design efforts integrating VO$_2$ into more sophisticated RF architectures.

## II. <u>Experimental</u>

VO$_2$ was integrated into a 50 Ω coplanar waveguide (CPW) transmission line by micro-fabrication, as described below. This configuration simplifies fabrication for RF devices because it eliminates the need for back side metallization or grounding vias. The devices were designed to interface with 3-terminal, ground-signal-ground (GSG) microwave probes and are arranged in a 2-port series configuration.

VO$_2$ films were grown by RF magnetron sputtering in an Ar/O$_2$ mixture onto single crystal (0001) sapphire substrates. Substrate temperature during growth was held at 550 °C and background pressure was 5 mTorr. The films, about 120 nm thick, were patterned by photolithography and reactive-ion etched (10/5 sccm Ar/CF$_4$, 78 mTorr total pressure) into mesas with lateral dimensions of 50 μm x 25-35 μm. Ground and signal contacts of 20 nm Ti/200 nm Au were patterned using e-beam evaporation and lift-off. The final dimensions of devices in this work were $W$ = 25 μm and $L$ = 5-15 μm. An optical micrograph top view of a representative device is shown in Fig. 1a. A side view schematic of the signal trace is shown in Fig. 1b.

DC *I-V* measurements were performed using a Keithley 2400 sourcemeter on a hot chuck probe station capable of 5 – 200 °C temperature control. Resistance was extracted by taking the inverse slope of *I-V* measurements in the linear low voltage regime with voltage swept between ± 0.1 V. Device switching speed was characterized at room temperature in a separate system in a configuration as shown in Fig. 3a. A continuous wave (CW) source (Agilent N5181A) of 2.3 GHz and 0 dBm was incident on the $VO_2$ CPW and the output was monitored by an oscilloscope (Tektronix DPO 72004). The CW source and oscilloscope used provided switching speed resolution below 0.5 ns. The $VO_2$ devices were switched to the metallic phase during transient measurements with a parallel DC sourcemeter, which was isolated from the CW supply using DC blocks. Output RF power was measured in the same setup with an Agilent E4417A power meter and E4413A power head in place of the oscilloscope. *S*-parameters were measured in a configuration as shown in Fig. 4a using an Agilent N5241A vector network analyzer (VNA) with a -5 dBm signal. An *S*-parameter calibration procedure was used to place the measurement reference plane at the end of the GSG probe tips. 150 µm GSG probes (GGB 40A) were used, and a Short, Open, Load, Through (SOLT) calibration was performed utilizing the GGB CS-5 calibration substrate. This substrate includes metal patterns and calibration coefficients (loaded into the VNA) to measure the response of a Short, Open, Load and Through which allows for the de-embedding of test setup losses and parasitics. The calibration is controlled via the VNA, which steps through each calibration standard.

**III. Results and discussion**

### A. DC characterization

A representative resistance-temperature (*R-T*) measurement of a VO$_2$ device studied in this work is given in Fig. 2a. The resistance changes by three orders of magnitude from room temperature to the metallic phase, with the phase transition occurring at 74 °C on heating and 67 °C on cooling, as determined by the peak in d(ln *R*)/d*T*. Along with thermal excitation, the MIT of VO$_2$ can also be triggered electrically in *I-V* measurements. To compare with the *R-T* curve, in Fig. 2b we plot data from an *I-V* measurement in the form of static resistance, *V/I*, as a function of voltage. For these measurements, we connected a series resistor (50 – 500 Ω) to prevent device failure due overheating and then subtracted the measured series resistance from the *V/I* data. The hysteresis in the *V/I-V* plot is similar to that observed in *R-T* and is indicative of an electrically-driven phase transition. We will discuss the E-MIT characteristics in greater depth below.

A notable aspect of the *I-V* curves measured on our devices is that the electrically-driven resistance change ($\Delta R_E$ = *R*(Off)/*R*(On)) is about three orders of magnitude across the E-MIT, essentially equivalent to the thermally-driven resistance change ($\Delta R_T$) measured in *R-T* from room temperature to 100 °C. The $\Delta R_E$ observed is significantly larger than most literature reports,[2-8] as noted in the introduction, and the comparable $\Delta R_E$ and $\Delta R_T$ values suggest that the full VO$_2$ device volume is being switched. In the above cited references, film thicknesses and metrology for determining $\Delta R_E$ are similar to our study, and thus it is reasonable to make a direct comparison. In a temperature-driven MIT, the temperature of the whole device chip is controlled by thermal contact with a hot chuck, and therefore the entire volume of the VO$_2$ device is switched to the metallic phase at temperature sufficiently above the MIT temperature.[17] The observation of low $\Delta R_E$ may hence imply incomplete switching of the VO$_2$ film, possibly due to non-uniformities in film composition or crystallographic orientation. For example, if only (nano)filaments or interfacial layers switch to the metallic phase in *I-V* in the above references, this will shunt the remainder of the film, leading to an E-MIT with an On state resistance value larger than if the entire film was switched. We suggest that because $\Delta R_E \approx \Delta R_T$ for the CPW $VO_2$ devices here, we may be observing electrical switching of the full $VO_2$ film volume. Note that large metallic domains in E-MIT on the surface of $VO_2$ have been previously reported,[18] while our results imply complete switching in the bulk of the film.

### B. Transient analysis of electrical switching

The transient response of the E-MIT switch is examined in Fig. 3. As drawn schematically in Fig. 3a, a 2.3 GHz CW source is incident on the $VO_2$ CPW and the output is monitored on an oscilloscope while the device is electrically switched with an isolated DC supply. Note that the high frequency CW periodicity is not apparent on these time scales and that the envelope of the output contains the relevant transient information. This method is preferred for high frequency RF switching studies. A representative *I-V* curve of the device for which the transient response is examined is shown in Fig. 3b. Here, we manually limit the compliance current ($I_{limit}$) on the sourcemeter to 100 mA to prevent overheating instead of using a series resistor. This eliminates artifacts due to high frequency reflections from added circuit components. It can be seen for this device that, as voltage is increased, there are three small abrupt jumps in current before a much larger abrupt jump at ~3 V to the compliance limit. The transient response is measured during such an *I-V* measurement. As seen in Fig. 3c, when the DC source is off ($t < -15$ μs), the measured output envelope is low. The DC source is a standard sourcemeter and does not provide a fast ramp rate when triggered. On the time scales studied here, triggering the DC source is effectively equivalent to applying a triangular input pulse. As the triangle input approaches ~3V, the output envelope abruptly increases in amplitude, indicating that CW transmission is significantly enhanced because the $VO_2$ channel is switched to the metallic phase. After switching, the DC supply drops to a steady state voltage that reflects the compliance current and on-state resistance of the $VO_2$ device.

It can be seen that as the DC supply is ramped, there are actually several small jumps in output envelope voltage before a large jump at ~21.4 μs, as outlined in the figure in red. These small jumps reflect the similar small jumps observed in *I-V* in Fig. 3b, denoting that we are directly probing the E-MIT transient measured by *I-V*. Note that the DC source voltage does not begin to drop (current does not go to $I_{limit}$) until the last large jump occurs, suggesting that this last jump corresponds to full switching of the $VO_2$. There are damped oscillations in the output envelope after the large jump and before steady state is reached, indicative of an underdamped second-order system. Oscillations associated with E-MIT switching have been observed before in $VO_2$ measurements and have been attributed to capacitive effects in the film with coexisting metallic and insulating phases.[8,17,19,20] The oscillation period in some of those works is of the same order as observed here, suggesting a similar mechanism in both cases, although there may also be capacitance effects from the test circuitry.

Rise time (Δ*t*) for a positive voltage step is generally defined as the difference between the time at which the signal is at 10% above its Off value to the time at which it is at 10% below its On value, *i.e.* Δ*t* = |$t_{off}$ − $t_{on}$| = |$t(V = 1.1V_{off})$ − $t(V = 0.9V_{on})$|. However, there are several effects in the transient response of the $VO_2$ E-MIT that suggest varying values of Δ*t*. We focus here only on the final large jump because the time between the small jumps is controlled by the DC supply ramp rate and is not a fundamental film property with this test setup. From an RF device perspective, a switch is not considered On until the transient reaches steady state.

Furthermore, $t_{on}$ is more stringently defined as the time at which $V \geq 0.9V_{on}$ for all $t \geq t_{on}$, which accounts for oscillations in which $V$ may temporarily decrease below $0.9V_{on}$ before reaching steady state. Within this definition, we find that the steady state rise time averaged between upper and lower envelopes is $\Delta t_{ss} = 5.13 \pm 0.05$ μs. Besides the RF device definition and perhaps more intrinsically, the VO$_2$ device can also be considered On at the peak of the first oscillation after the last abrupt voltage jump (see Fig. 3d). In this case, the switch is regarded as On before the oscillations dampen out. This is the definition typically found in the VO$_2$ E-MIT literature,[2,8,9] although it may not be as relevant as the former definition for electronic device design. With this definition based solely on the voltage jump, we have $\Delta t_j = 14.1 \pm 2.7$ ns, of the same order as quoted in the literature cited above considering differences in device dimensions.

There is much published work discussing whether the VO$_2$ E-MIT mechanism is due to Joule heating, electric field, high-level carrier injection, or some combination of these effects.[1-3,7,21-23] For the Joule heating mechanism, a simple 1D heat diffusion model for $I$-$V$ measurements on a two-terminal thin film device can be made, giving

$$\kappa \frac{\partial^2 T}{\partial x^2} - \gamma T + \frac{IV}{\Omega} = \rho C_p \frac{\partial T}{\partial t} \qquad (1)$$

where $T$ is the excess temperature of the film above ambient, $\kappa$ is the film thermal conductivity, $\gamma = \kappa_{sub}/d_{film}d_{sub}$ is the thermal coupling coefficient with the substrate,[24] $I \cdot V$ is the applied electrical power, $\Omega$ is the device volume, $\rho$ is the film mass density, and $C_p$ is the specific heat. This simple equation assumes perfect thermal contact (zero thermal resistance) at the contact-film and film-substrate interfaces and it accounts for heat lost through the substrate. With respect to VO$_2$, the minimum switching time ($\Delta t_{min}$) to achieve a temperature difference $\Delta T$ is usually calculated assuming no spatial temperature gradient and no heat loss through the substrate,[1,3] in which case the heat equation can be solved simply as

$$\Delta t_{\min} = \frac{\rho C_p \Omega \Delta T}{IV}. \tag{2}$$

If more complex heat loss mechanisms are taken into account, then the calculated $\Delta t$ will increase, which is why it is designated as the minimum switching time here. With the parameters $\rho$ = 4340 kg/m$^3$, $C_p$ = 690 J/kg·K, $\Omega$ = (5 μm x 25 μm x 150 nm), $\Delta T$ = 40 K, $I$ = 6 x 10$^{-4}$ A, and $V$ = 3 V, we calculate $\Delta t_{\min}$ = 1.2 μs. The calculated $\Delta t_{\min}$ here is orders of magnitude larger than the measured $\Delta t_j$, in agreement with various literature reports.[1,3,8,9] This is commonly taken as evidence that Joule heating alone cannot account for the fast intrinsic switching times in the E-MIT. However, in reported VO$_2$ devices where $\Delta R_E < \Delta R_T$, if filamentary or interfacial conduction is dominant, the volume used in Eq. 2 may be much lower than the actual device volume, which may account for the lower $\Delta t_j$ observed. Yet here we also observe $\Delta t_{\min} \gg \Delta t_j$ in our devices where $\Delta R_E \approx \Delta R_T$, further suggesting the role of a non-Joule heating mechanism for fast $\Delta t_j$ in E-MIT.

While the calculated switch time based on Joule heating, $\Delta t_{\min}$, is much larger than the fast rise time of the large voltage jump ($\Delta t_j$) in Fig. 3d, it is in rather good agreement with the rise time of the steady state signal ($\Delta t_{ss}$). It should be noted that in Ref. 8, abrupt VO$_2$ switching transients were studied and slowly damped oscillations after a fast ~5 ns initial jump are similarly observed. In that work, $\Delta t_{ss}$ (~75 ns) is of the same order as $\Delta t_{\min}$ (48-60 ns), as calculated for Joule heating in their respective device geometry, in agreement with our finding of $\Delta t_{\min} \sim \Delta t_{ss}$ here. If there is indeed a link between Joule heating and $\Delta t_{ss}$, it may provide evidence for a two-process mechanism of the E-MIT. Such a mechanism may involve some non-heating related effect such as an electric-field or carrier injection, which causes the initial fast current jump, followed by a temperature rise in response to the sudden increase in dissipated power, which causes the slow dampening to steady state.

## C. RF characterization

The room temperature RF characteristics of a $VO_2$ CPW were examined with the test configuration shown in Fig. 4a. The $VO_2$ device was switched in a similar manner as the transient measurements and return loss ($|S_{11}|$) and insertion loss ($|S_{21}|$) were recorded in Off and On states. Fig. 4b demonstrates how the insertion loss at 10 GHz varies as a function of compliance current. With $I_{limit}$ = 10 mA, the device is turned on, as indicated by the reduction in insertion loss. Little improvement in insertion loss occurs for $I_{limit}$ higher than 70-80 mA and therefore we choose this range as a set point for On state characterization. As seen in Fig. 4c, in the On state, the insertion loss of the device is extremely flat as a function of frequency and has a value of 2.95 dB at 13 GHz. In addition, the device is fairly well matched at about -11 dB (Fig. 4d) across the frequency range. This indicates mismatch losses of 0.36 dB, resulting in dissipative losses of 2.6 dB. A high frequency structural simulator (HFSS) model of the device shows that a simple geometric modification can further reduce insertion loss in the On state, as discussed below.

In the Off state, isolation is set primarily by the signal trace geometry. Since the $VO_2$ conductivity is significantly lower in the Off state as opposed to the On state, signal coupling occurs due to the parasitic capacitance formed by the gap between electrodes in the signal trace. Insertion loss measured better than 25 dB at 13 GHz. Capacitive coupling is reduced at lower frequencies, leading to greater insertion loss. The input match, $S_{11}$ and $S_{22}$, is high, indicating no power is dissipated by the device in the Off state and that incident power is well reflected off of the device. The On and Off state insertion loss results are in good agreement with published reports.[3,5]

Plotting $S_{11}$ on a Smith chart (Fig. 5a) shows that the $VO_2$ switch behaves like a variable resistor by traveling along the real axis as bias current increases. In the Off state, the impedance locus is at the right edge of the Smith chart, as expected for an open circuit. The locus rotates clockwise with frequency due to the parasitic capacitance formed by the gap between contacts in the signal trace. As bias current increases, $S_{11}$ travels along the real axis towards the center of the Smith chart. At 70 mA bias current, the switch conductivity shorts the parasitic capacitance. The variable resistor behavior likely originates from partial switching of the channel at low current, i.e. $R_{Total} = t \cdot R_{On} + (1 - t) \cdot R_{Off}$, where $t$ is the fraction of switched material at a given bias current.

An important characteristic for RF devices is the output RF power response ($P_{out}$) as a function of input RF power ($P_{in}$). In Fig. 5c, we plot this characteristic in the On state for a 2.3 GHz input signal and various levels of bias current. Up to around 0.5 W (~27 dBm) the $VO_2$ response is linear with no observable compression. At high $P_{in}$ the output power becomes slightly superlinear ("expansion") due to reduced insertion loss, which is more clearly shown in Fig 5d. The devices were able to operate reliably for power levels between 0.25-0.5 W (~24-27 dBm), with some devices able to handle as much as 2 W (~33 dBm) of continuous forward power. The "expansion" at high $P_{in}$ is much more noticeable for the device switched with 25 mA compliance current, and it is due to the $VO_2$ not being fully switched On at relatively low DC bias levels. The added RF power is sufficient to compensate for low DC bias and drive the $VO_2$ further into the metallic phase, which results in a reduction of insertion loss with increasing $P_{in}$. The decrease in insertion loss indicates that more RF signal reaches the output terminal due to lower resistance of the CPW. This is similar to decreasing insertion loss with increasing bias current (*i.e.* higher metallicity, lower resistance) as in Fig. 4b. The abruptness of the insertion loss drop in Fig. 5d also resembles the abruptness of the E-MIT in Fig. 3b. These observations suggest that at sufficiently high power levels, electromagnetic radiation behaves similarly as DC electrical excitation and that RF power may be able to trigger the MIT of $VO_2$. This encouraging finding may indicate a conceptually new excitation source (RF power) for the $VO_2$ phase transition that could expand the possible application-space of $VO_2$ to encompass RF-sensitive circuitry such as RF receiver protectors.

### D. Modeling

For designing future $VO_2$ RF circuit components, it is necessary to be able to model frequency-dependent device performance. Initially, a lumped element circuit model was created to gain insight into device operation. Since the results of this model correspond only to a single device geometry, a 3D electromagnetic field model was created using the results of the lumped element model. The 3D model allows for future device design by giving accurate predictions of RF performance for different device geometries.

Analyzing the measured *S*-parameters on a Smith chart for different device bias conditions gives an intuitive understanding of device behavior and guides topology selection for the lumped element model. The Smith chart behavior is characteristic of a simple parallel *RC* circuit that has a current-dependent resistance, as illustrated in Fig. 5b. Using this circuit topology as a starting point, the *S*-parameter Simulator in Agilent's Advanced Design System (ADS) software was used to fit this circuit model to the measured data. The ADS *S*-parameter Simulator is a frequency-domain circuit simulator that analyzes RF circuit performance under small signal AC drive conditions. Resistor and capacitor values are used as fitting parameters to fit the circuit response to the measured *S*-parameters. Since $S_{11}$ is equal to the reflection coefficient of a 2-port measurement, the impedance looking into the device can be obtained using the relationship $Z_1 = Z_0(1 + S_{11})/(1 - S_{11})$, where $Z_1$ is the input impedance of port 1 and $Z_0$

= 50 Ω is the reference impedance. Figs. 6a-b show the results of the ADS lumped element model (red dashed line) for $S_{21}$ and $S_{11}$ under 70 mA bias current. Good agreement with measured data (black solid line) is obtained with an On state resistance of 42 Ω and an Off state resistance of 51 kΩ. The extracted parasitic capacitance is approximately 0.006 pF. The extracted resistance change magnitude between On and Off states agrees well with the E-MIT *I-V* data. The fitting of the *RC* lumped element frequency response to the measured *S*-parameter data validates the lumped element model in this frequency range.

While the variable resistance can be simply understood on the basis of the MIT, the parallel capacitor arises from the proximity coupling of the two signal lines and the dielectric nature of the insulating phase. When the two-terminal device is on, the low impedance of the $VO_2$ channel effectively shorts out the parasitic capacitance and dominates the input impedance. This value is small as expected from the geometry of the two input lines and is the reason the RF response is significantly broadband. In the On state, the resistance is low enough such that reflected RF power is minimized. In the Off state the device resistance is 51 kΩ, which provides a very large reflection coefficient and rejects incident RF energy similar to an open circuit.

In addition to the lumped element model, a 3D electromagnetic model of the device was created to be used as a tool for designing devices of different geometries. Ansoft High Frequency Structure Simulator (HFSS) was used for the creation of this model. HFSS is a 3D finite element solver that can generate *S*-parameters from the calculated electromagnetic field pattern inside a structure. A 3D model of the device was created in HFSS and software-default material properties were assigned to the sapphire substrate and Au metallization. Conductivity for the $VO_2$ channel was calculated based on the results of the lumped element model and the device geometry. On state conductivity was calculated to be 83,000 S/m and was entered into the HFSS simulation. Results of the HFSS model for the On state are given in Figs. 6c-d (red dashed line), showing good agreement with experimental data (black solid line). In addition, by resizing the HFSS model device geometry to 75 μm x 10 μm junctions to reduce the On state resistance, the insertion loss is projected to be lowered to less than 1 dB with broadband characteristics up to 40 GHz (green dashed line).

The assumption of parasitic capacitance arising from coupling between signal electrodes and the dielectric nature of $VO_2$ was examined by studying devices of different geometry. Insertion loss in the Off state of CPW devices of length 5 μm and 15 μm was measured and simulated using HFSS, as shown in Fig. 6e. Within this model of the parasitic capacitance, it is expected that the longer device should have a larger insertion loss in the Off state due to smaller capacitance (higher impedance). The HFSS simulation predicts a 2.9 dB increase in $|S_{21}|$ in the Off state when the $VO_2$ channel length is increased from 5 to 15 μm. Measured S-parameters also show a 3.2 dB increase in $|S_{21}|$ as the device channel is increased from 5 to 15 μm. The same trend observed in both the simulated and measured data supports the proposed small signal lumped element model.

## IV. Conclusions

Electrical switching and microwave RF properties of two-terminal $VO_2$ coplanar waveguides were examined. A large resistance change in the electrically-driven phase transition of the same order ($>10^3$) as in the temperature-driven phase transition was observed. Two distinguishable time scales for electrical switching were observed, one in accordance with Joule heating and one much faster, which may suggest a two-process mechanism for the electrical switching. RF power response showed that increasing input power lowers the On state resistance of VO$_2$ similarly as increasing bias current. The RF response showed flat insertion loss at 2.95 dB up to 13.5 GHz in the On state, increasing to over 25 dB in the Off state. The RF properties were modeled using a parallel resistor-capacitor lumped circuit as well as 3D electromagnetic simulations. The results could be of relevance to designing RF circuits utilizing phase transition oxide materials.

**Acknowledgments**

This work was supported by ONR N00014-12-1-0451. The authors acknowledge Doug White, Reed Irion, and Glenn Thoren for technical discussions.

**Figure captions**

**Figure 1:** **(A)** Optical micrograph of coplanar waveguide studied in this work. **(B)** Side view schematic of signal trace.

**Figure 2:** **(A)** Representative *R-T* curve of $VO_2$ device studied in this work demonstrating thermally-driven MIT. Resistance normalized to room temperature value. **(B)** Static resistance as a function of voltage for similar $VO_2$ device as in (a) showing large change in resistance of same order of magnitude as *R-T* across E-MIT. Resistance normalized to low voltage value.

**Figure 3:** **(A)** Schematic of switching speed test setup. **(B)** *I-V* curve of transient switching event examined in panels (c-d). **(C)** Transient output of DC source and $VO_2$ device showing switching of $VO_2$ at a critical DC voltage. Voltage signal from $VO_2$ is scaled by 5x for clarity. **(D)** Zoomed-in view of large, final switching event with smoothed data (blue) to more easily distinguish rise time.

**Figure 4:** **(A)** Schematic of RF test setup. **(B)** Insertion loss at 10 GHz of $VO_2$ device vs. maximum bias current showing large change with increasing current. **(C-D)** $S_{21}$ and $S_{11}$ vs. frequency in Off state (black line) and On state (red line) at 70 mA bias.

**Figure 5:** **(A)** $S_{11}$ plotted on Smith chart for varying bias current levels for $f = 0.01 - 13.5$ GHz. **(B)** Lumped element model of $VO_2$ switch device deduced from Smith chart. **(C)** Output RF power as a function of input RF power for 2.3 GHz signal for varying bias

current levels. **(D)** Insertion loss ($=P_{in} - P_{out}$) as a function of input RF power extracted from (c) showing decreasing insertion loss at high input power, indicative of RF-triggering of phase transition.

**Figure 6:** **(A-B)** Comparison of $S_{21}$ and $S_{11}$ measured experimentally (black solid line) and simulated using ADS lumped element model (red dashed line) showing good agreement. **(C-D)** Additional modeling of $VO_2$ device using HFSS shows good agreement. Projected *S*-parameters of optimized geometry (green dashed line) show excellent broadband characteristics up to 40 GHz. **(E)** Off state $S_{21}$ measured (solid lines) and simulated (dashed lines) using HFSS for devices of length 5 μm (black lines) and 15 μm (red lines). Increase in insertion loss for 15 μm device is well modeled by decrease in parasitic capacitance.

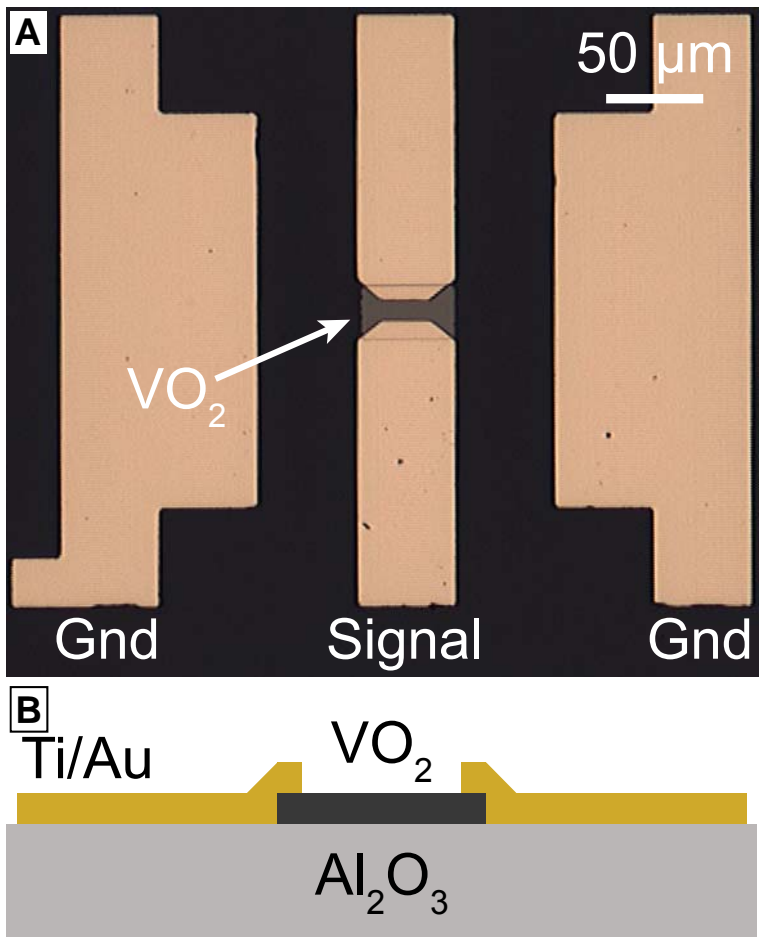

Figure 1

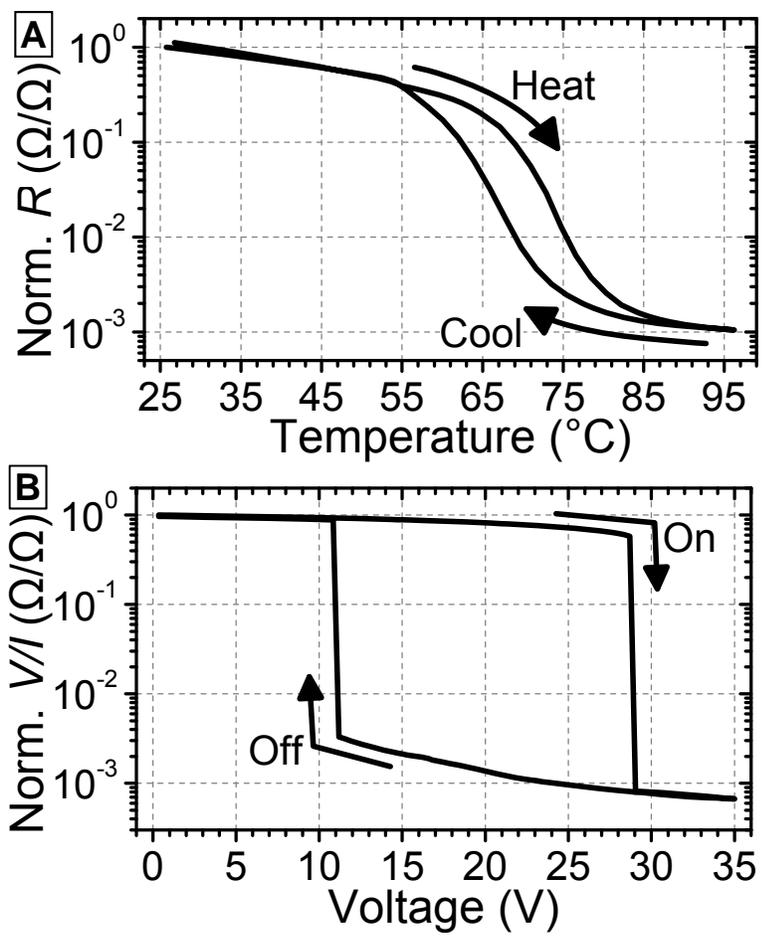

Figure 2

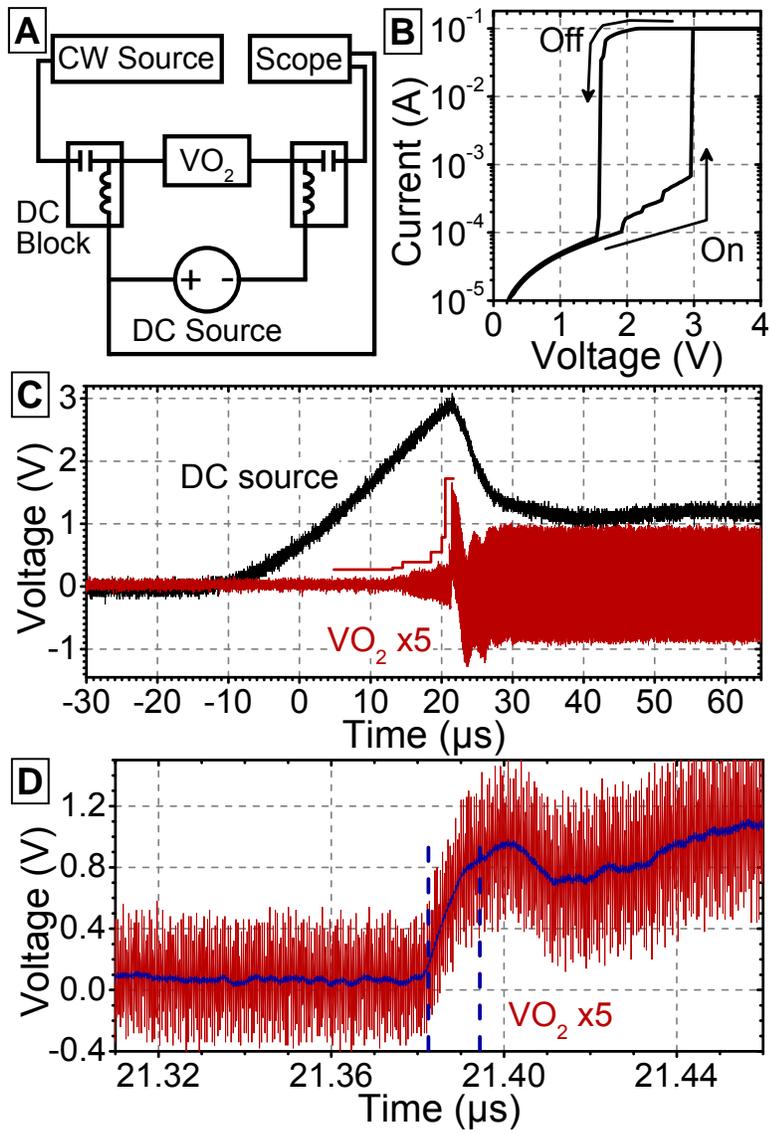

Figure 3

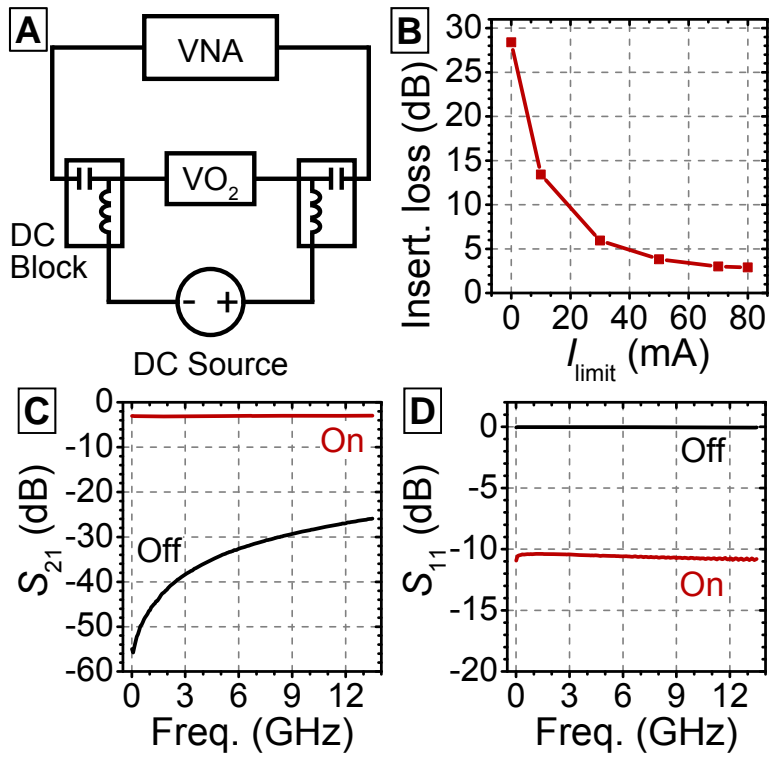

Figure 4

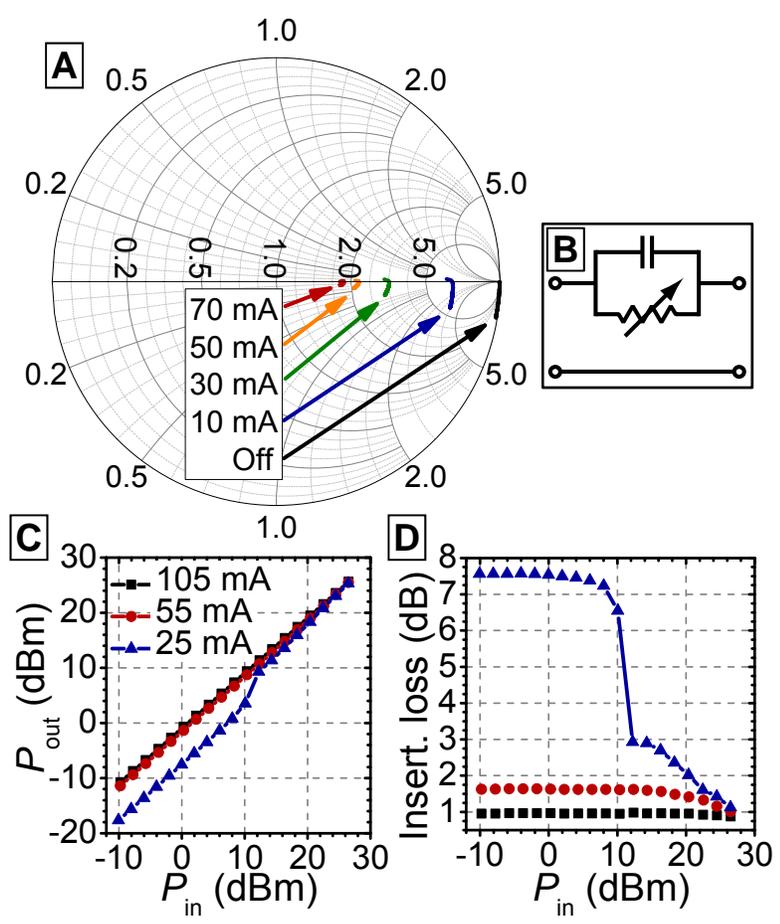

Figure 5

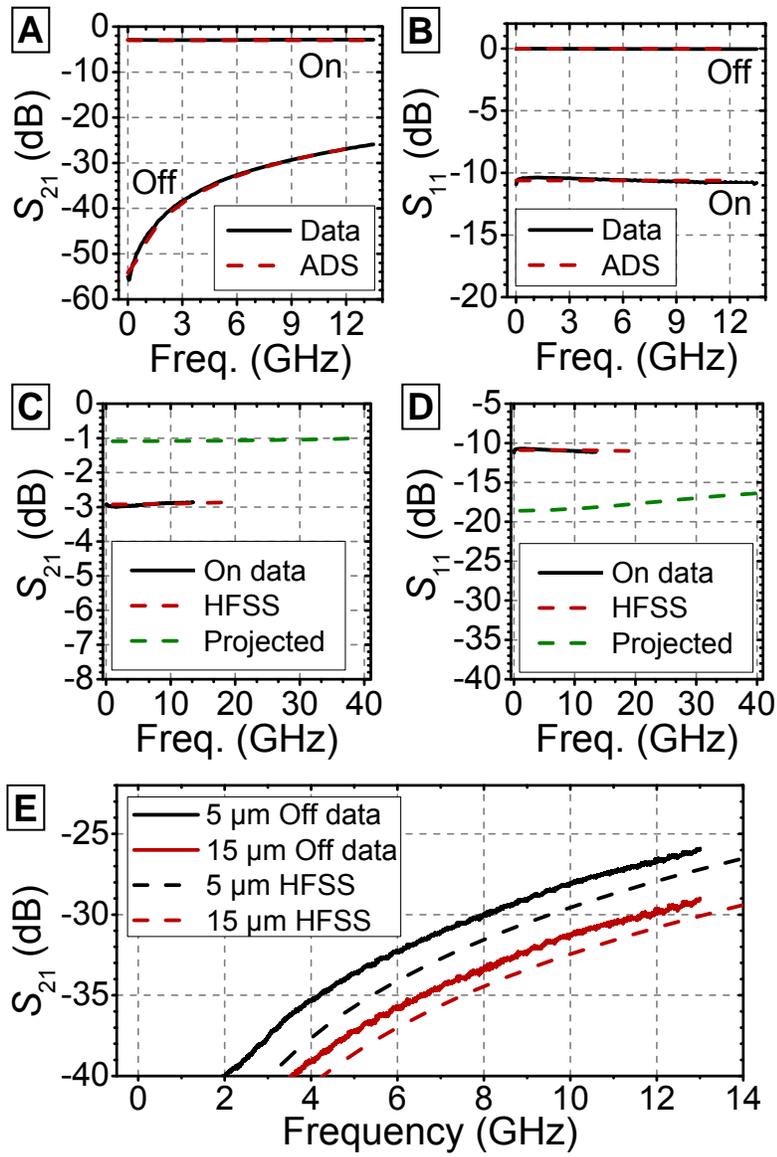

Figure 6